\documentclass[12pt,reqno]{amsart}

\usepackage{epsfig}
\usepackage{amsmath, amsthm, amsfonts}
\usepackage{graphicx}

\numberwithin{equation}{section}

\newcommand{\R}{{\mathbb R}}
\newcommand{\C}{{\mathbb C}}
\newcommand{\N}{{\mathbb N}}

\newcommand{\al}{\alpha}
\newcommand{\be}{\beta}
\newcommand{\ga}{\gamma}

\newcommand{\la}{\lambda}
\newcommand{\ep}{\varepsilon}
\newcommand{\de}{\delta}
\newcommand{\De}{\Delta}

\newcommand{\sg}{\sigma}

\newtheorem{theo}{{\sc \bf Theorem}}[section]

\newtheorem{prop}[theo]{{\sc \bf Proposition}}

\begin{document}

\title[Exact solution of the six-vertex model]
{Exact solution of the six-vertex model with domain wall boundary conditions.
Ferroelectric phase}

\author{Pavel Bleher}
\address{Department of Mathematical Sciences,
Indiana University-Purdue University Indianapolis,
402 N. Blackford St., Indianapolis, IN 46202, U.S.A.}
\email{bleher@math.iupui.edu}

\author{Karl Liechty}
\address{Department of Mathematical Sciences,
Indiana University-Purdue University Indianapolis,
402 N. Blackford St., Indianapolis, IN 46202, U.S.A.}
\email{kliechty@math.iupui.edu}

\thanks{The first author is supported in part
by the National Science Foundation (NSF) Grant DMS-0652005.}

\date{\today}

\begin{abstract} This is a continuation of the paper \cite{BF} of Bleher and Fokin,
in which the large $n$ asymptotics is obtained for the partition function $Z_n$ of the
six-vertex model with domain wall boundary conditions in the disordered phase. In the present paper
we obtain the large $n$ asymptotics of $Z_n$
in the ferroelectric phase. We prove that for any $\ep>0$, as $n\to\infty$,
$Z_n=CG^nF^{n^2}[1+O(e^{-n^{1-\ep}})]$, and we find the exact values of the constants $C,G$ and $F$.
The proof is based on the large $n$ asymptotics for the underlying discrete orthogonal polynomials and 
on the Toda equation for the tau-function.
\end{abstract}

\maketitle

\section{Introduction and formulation of the main result}

\subsection{Definition of the model}
The six-vertex model, or the model of two-dimensional ice, is stated on a square $n\times n$
lattice with arrows on edges. The arrows obey the rule that at every vertex there 
are two arrows 
pointing in and two arrows pointing out. Such rule is sometimes 
called the {\it ice-rule}. There are only six possible configurations of arrows at each 
vertex, hence the name of the model, see Fig.~1. 

\begin{center}
 \begin{figure}[h]\label{arrows}
\begin{center}
   \scalebox{0.52}{\includegraphics{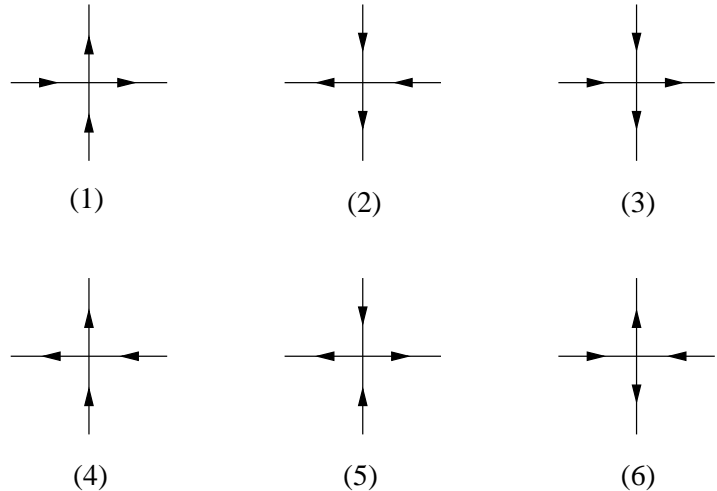}}
\end{center}
        \caption[The six arrow configurations allowed at a vertex]{The six arrow configurations allowed at a vertex.}
    \end{figure}
\end{center}

We will consider the {\it domain wall boundary conditions} (DWBC), 
in which the arrows on the upper and lower boundaries point in the square, 
and the ones on the left and right boundaries point out. 
One possible configuration with DWBC on the $4\times 4$ lattice is shown on Fig.~2.
\begin{center}
 \begin{figure}[h]\label{DWBC}
\begin{center}
   \scalebox{0.52}{\includegraphics{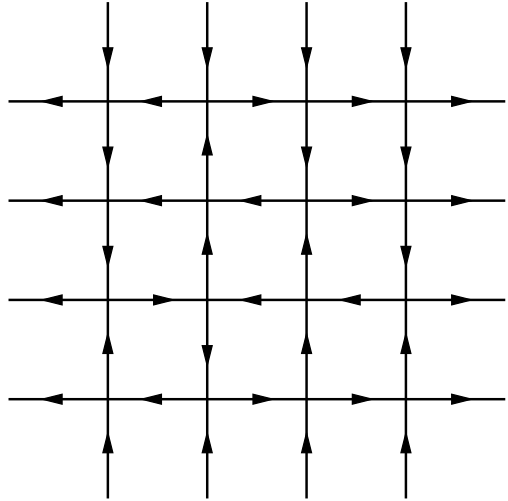}}
\end{center}
        \caption[An example of $4\times4$ configuration]
{An example of $4\times4$ configuration with DWBC.}
    \end{figure}
\end{center}

For each possible vertex state we assign a weight $w_i,\; i=1,\dots,6$, 
and define, as usual, the partition function as a sum over all possible 
arrow configurations of the product of the vertex weights,
\begin{equation}\label{lattice_11}
Z_n=\sum_{{\rm arrow\; configurations}\;\sigma}w(\sigma),
\qquad w(\sigma)=\prod_{x\in V_n} w_{t(x;\sg)}=\prod_{i=1}^6w_i^{N_i(\sigma)},
\end{equation}
where $V_n$ is the $n\times n$ set of vertices,
$t(x;\sg)\in\{1,\ldots,6\}$ is the type of configuration $\sg$
at vertex $x$ according to Fig. 1, and
$N_i(\sigma)$ is the number of vertices of  type $i$ in  
the configuration $\sg$. The sum is taken over all possible configurations
obeying the given boundary condition. The Gibbs measure is defined then
as
\begin{equation}\label{lattice_12}
\mu_n(\sg)=\frac{w(\sg)}{Z_n}\,.
\end{equation}
Our main goal is to obtain the large $n$ asymptotics of the partition function $Z_n$.

 The six-vertex model has six parameters: the weights $w_i$. By using some conservation
laws it can be reduced to only two parameters. It is convenient to derive
the  conservation laws from the {\it height function}.

\subsection{Height function}
Consider the
dual lattice,
\begin{equation}\label{hf_1}
V'=\{x=(i+\frac{1}{2},j+\frac{1}{2}),\quad  0\le i,j\le n\}.
\end{equation}
Given a configuration $\sg$ on $E$,
an integer-valued function $h=h_\sg$ on $V'$ is called a {\it height function}
of $\sg$,
if for any two neighboring points $x,y\in V'$, $|x-y|=1$, we have that
\begin{equation}\label{hf_2}
h(y)-h(x)=(-1)^s,
\end{equation}
where $s=0$ if the arrow $\sg_e$ on the edge $e\in E$, 
crossing the segment $[x,y]$, is oriented in such a way that
it points from left to right with respect to the vector $\vec{xy}$ , and $s=1$ if $\sg_e$ is oriented from right to left with respect to $\vec{xy}$. 
The ice-rule ensures that the height function $h=h_\sg$ exists for any 
configuration $\sg$. It is unique up to addition of a constant.
Fig. 3 shows a $5\times 5$ configuration with a height function, and
the corresponding alternating sign matrix,
which is obtained from the configuration by replacing the vertex (5) of Fig.~1 by 1,
the vertex (6) by $(-1)$, and all the other vertices by 0.
\begin{center}
 \begin{figure}[h]\label{asm[1]}
\begin{center}
   \scalebox{0.7}{\includegraphics{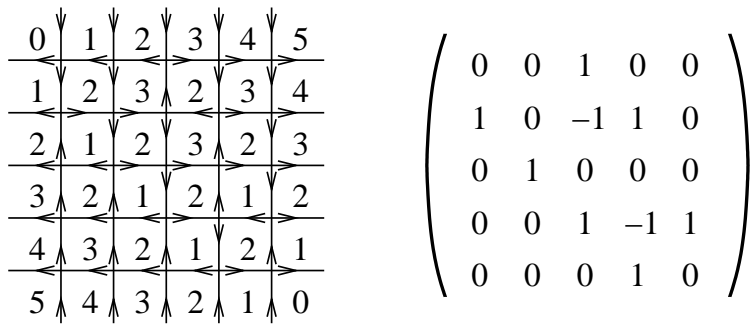}}
\end{center}
        \caption[A $5\times 5$ configuration with a height function, and
the corresponding ASM. ]
{A $5\times 5$ configuration with a height function and
the corresponding alternating sign matrix. }
    \end{figure}
\end{center}
 Observe that if $h_1(x),h_2(x),h_3(x),h_4(x)$ are the four values 
of the height function around a vertex $x=(j,k)$, enumerated in the positive direction
around $x$ starting from the first quadrant, 
then the value of the element $a_{jk}$ of the ASM is equal to
\begin{equation}\label{asm_3}
a_{jk}=\frac{h_1(x)-h_2(x)+h_3(x)-h_4(x)}{2}\,.
\end{equation}

\subsection{Conservation laws} Conservation laws are obtained in the paper \cite{FS}
of Ferrari and Spohn, as a corollary of a path representation of the six-vertex model.
We will derive them from the height function representation.
Consider the height function $h=h_\sg$ on a diagonal sequence of points defined by the formula,
\begin{equation}\label{cl_1}
x_j=x_0+(j,j),\quad 0\le j\le k,
\end{equation}
where both $x_0$ and $x_k$ lie on the boundary $B'$ of the dual lattice $V'$,
\begin{equation}\label{cl_2}
\begin{aligned}
B'&=\{ x=(i+\frac{1}{2},\frac{1}{2}),\;0\le i\le n\}
\cup \{ x=(m+\frac{1}{2},j+\frac{1}{2}),\;0\le j\le n\}\\
&\cup \{ x=(i+\frac{1}{2},n+\frac{1}{2}),\;0\le i\le n\}
\cup \{ x=(\frac{1}{2},j+\frac{1}{2}),\;0\le j\le n\}.
\end{aligned}
\end{equation}
Then it follows from the definition of the height function, that
\begin{equation}\label{cl_3}
h(x_j)-h(x_{j-1})=
\left\{
\begin{aligned}
&2,\quad{\rm if}\quad t(x;\sg)=3,\\
&-2,\quad{\rm if}\quad t(x;\sg)=4,\\
&0,\quad{\rm if}\quad t(x;\sg)=1,2,5,6,
\end{aligned}
\right.
\end{equation}
where
\begin{equation}\label{cl_4}
x=\frac{x_j+x_{j-1}}{2}\,.
\end{equation} 
Hence
\begin{equation}\label{cl_5}
0=h(x_k)-h(x_0)=2N_3(\sg;L)-2N_4(\sg;L),
\end{equation}
where $N_i(\sg;L)$ is the number of vertex states of type $i$ in $\sg$
on the line
\begin{equation}\label{cl_6}
L=\{x=x_0+(t,t),\;t\in\R\}.
\end{equation}
The line $L$ is parallel to the diagonal $y=x$. By summing up over
all possible lines $L$, we obtain that
\begin{equation}\label{cl_7}
N_3(\sg)-N_4(\sg)=0,
\end{equation}
where $N_i(\sg)$ is the total number of vertex states of the type $i$
in the configuration $\sg$. 

Similarly, by considering lines $L$ parallel to the diagonal $y=-x$,
we obtain that
\begin{equation}\label{cl_8}
N_1(\sg)-N_2(\sg)=0.
\end{equation}
Also,
\begin{equation}\label{cl_9}
N_5(\sg)-N_6(\sg)=n,
\end{equation}
which follows if we consider lines $L$ parallel to the $x$-axis.

The conservation laws allow to reduce the weights $w_1,\ldots,w_6$
to 3 parameters. Namely, we have that
\begin{equation}\label{cl_11}
w_1^{N_1}w_2^{N_2}w_3^{N_3}w_4^{N_4}w_5^{N_5}w_6^{N_6}=
C(n)a^{N_1}a^{N_2}b^{N_3}b^{N_4}c^{N_5}c^{N_6},
\end{equation}
where
\begin{equation}\label{cl_12}
a=\sqrt{w_1w_2},\quad b=\sqrt{w_3w_4},\quad c=\sqrt{w_5w_6},
\end{equation}
and the constant
\begin{equation}\label{cl_13}
C(n)=
\left(\frac{w_5}{w_6}\right)^{\frac{n}{2}}\,.
\end{equation}
This implies the relation
between the partition functions,
\begin{equation}\label{cl_14}
Z_n(w_1,w_2,w_3,w_4,w_5,w_6)=C(n)Z_n(a,a,b,b,c,c),
\end{equation}
and between the Gibbs measures,
\begin{equation}\label{cl_15}
\mu_n(\sg;w_1,w_2,w_3,w_4,w_5,w_6)=\mu_n(\sg;a,a,b,b,c,c).
\end{equation}
Therefore, for fixed boundary conditions, like DWBC, the general weights
are reduced to the case when
\begin{equation}\label{cl_16}
w_1=w_2=a,\quad w_3=w_4=b,\quad w_5=w_6=c.
\end{equation}
Furthermore, 
\begin{equation}\label{cl_17}
Z_n(a,a,b,b,c,c)=c^{n^2}Z_n\left(\frac{a}{c},\frac{a}{c},\frac{b}{c},\frac{b}{c},1,1\right)
\end{equation}
and
\begin{equation}\label{cl_18}
\mu_n(\sg;a,a,b,b,c,c)=\mu_n\left(\sg;\frac{a}{c},\frac{a}{c},\frac{b}{c},\frac{b}{c},1,1\right),
\end{equation}
so that a general weight reduces to the two parameters, $\frac{a}{c},\frac{b}{c}\,$.

\subsection {Exact solution of the six-vertex model for a finite $n$}
Introduce the parameter
\begin{equation}\label{pf1}
\Delta=\frac{a^2+b^2-c^2}{2ab}\,.
\end{equation}
There are three physical phases in the six-vertex model: the ferroelectric phase, $\Delta > 1$; the anti-ferroelectric phase, $\Delta<-1$; and, the disordered phase, $-1<\Delta<1$. 
In the three phases we parametrize the weights in the standard way:
for the ferroelectric phase,
\begin{equation}\label{pf4}
a=\sinh(t-\ga), \quad
b=\sinh(t+\ga), \quad
c=\sinh(2|\ga|), \quad
0<|\ga|<t,
\end{equation}
for the anti-ferroelectric phase,
\begin{equation}\label{pf5}
a=\sinh(\ga-t), \quad
b=\sinh(\ga+t), \quad
c=\sinh(2\ga), \quad
|t|<\ga,
\end{equation}
and for the disordered phase
\begin{equation}\label{pf6}
a=\sin(\ga-t), \quad
b=\sin(\ga+t), \quad
c=\sin(2\ga), \quad
|t|<\ga.
\end{equation}
The phase diagram of the six-vertex model is shown on Fig.~4.
\begin{center}
 \begin{figure}[h]\label{PhaseDiagram1}
\begin{center}
   \scalebox{0.5}{\includegraphics{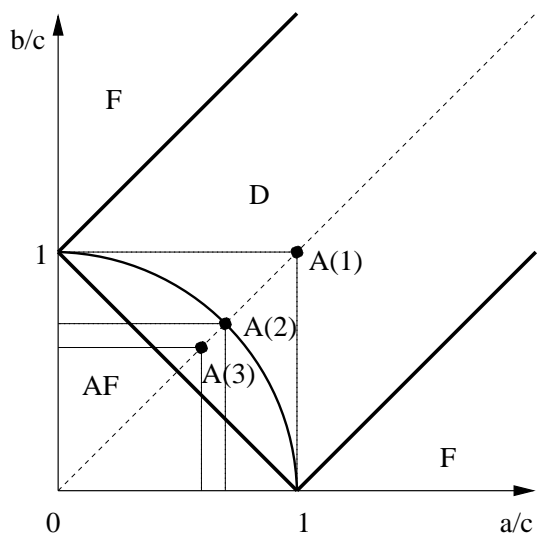}}
\end{center}
        \caption[The phase diagram of the model]{The phase diagram of the model, where {\bf F}, {\bf AF} and {\bf D} mark ferroelectric, antiferroelectric,  and disordered  phases, respectively. The circular arc corresponds to the so-called "free fermion" line, when $\Delta=0$, and the three
dots correspond to 1-, 2-, and 3-enumeration of alternating sign matrices.}
    \end{figure}
\end{center}
The phase diagram  and the Bethe-Ansatz solution  of the six-vertex model for periodic and anti-periodic
boundary conditions are thoroughly
discussed in the works of Lieb \cite{Lieb1}-\cite{Lieb4}, Lieb, Wu \cite{LW},
Sutherland \cite{Sut}, Baxter \cite{Bax}, Batchelor, Baxter, O'Rourke, Yung \cite{BBOY}.
See also the work of Wu, Lin \cite{WL}, in which  the Pfaffian solution for the six-vertex
model with periodic boundary conditions is obtained on the free fermion line, $\Delta=0$.

As concerns the six-vertex model with DWBC, it is noticed by Kuperberg \cite{Kup}, 
that on the diagonal, 
\begin{equation}\label{pf2}
\frac{a}{c}=\frac{b}{c}=x,
\end{equation}
 the six-vertex model with DWBC is equivalent to the $s$-enumeration of alternating
sign matrices (ASM), in which the weight of each such matrix is equal to $s^{N_-}$, 
where $N_-$ is the number of $(-1)$'s in the matrix and $s=\frac{1}{x^2}$. The exact
solution for a finite $n$ is known for 1-, 2-, and 3-enumerations of ASMs, see the 
works by Kuperberg \cite {Kup} and Colomo-Pronko \cite{CP1} for a solution
based on the Izergin-Korepin formula. A fascinating story of the discovery 
of the ASM formula is presented in the book  \cite{Bre} of Bressoud.
On the free fermion line, $\ga=\frac{\pi}{4}$, the partition function of
the six-vertex model with DWBC has a very simple form: $Z_n=1$. For a nice
short proof of this formula see the work \cite{CP1} of Colomo-Pronko. 

Here we will discuss  the ferroelectric phase, and we will use parametrization (\ref{pf4}).
Without loss of generality we may assume that
\begin{equation}\label{pf6a}
\ga>0,
\end{equation}
which corresponds to the region, 
\begin{equation}\label{pf6b}
b>a+c.
\end{equation}
The parameter $\Delta$ in the ferroelectric phase reduces to
\begin{equation}\label{pf6c}
\Delta=\cosh(2\ga).
\end{equation}

 The six-vertex model with DWBC was introduced 
by Korepin in \cite{Kor}, who derived an important 
recursion relation for the partition function of the model.
This lead to a beautiful determinantal formula of Izergin \cite{Ize}
for the partition function with DWBC.
A detailed proof of this formula and its generalizations are given in
the paper of Izergin, Coker, and Korepin \cite{ICK}. When the weights are parameterized according to (\ref{pf4}), the formula of Izergin is
\begin{equation} \label{pf7}
Z_n=\frac{[\sinh(t-\ga)\sinh(t+\ga)]^{n^2}}{\left(
\prod_{j=0}^{n-1}j!\right)^2}\,\tau_n\,,
\end{equation}
where $\tau_n$ is the Hankel determinant,
\begin{equation} \label{pf8}
\tau_n=\det\left(\frac{d^{j+k-2}\phi}{dt^{j+k-2}}\right)_{1\le j,k\le n},
\end{equation} 
and
\begin{equation} \label{pf9}
\phi(t)=\frac{\sinh(2\ga)}{\sinh(t+\ga)\sinh(t-\ga)}\,.
\end{equation}
An elegant derivation of the 
Isergin determinantal formula from the Yang-Baxter equation is 
given in the papers of Korepin, Zinn-Justin \cite{KZ} and Kuperberg \cite {Kup}
(see also the book of Bressoud \cite{Bre}).

One of the applications of the determinantal formula is that it
implies that 
the partition function $\tau_n$ solves the Toda equation
\begin{equation} \label{pf10}
\tau_n\tau''_n-{\tau'_n}^2=\tau_{n+1}\tau_{n-1},
\qquad n\ge 1,\qquad ({}')=\frac{\partial }{\partial t}\,,
\end{equation}
cf. the work of Sogo, \cite{Sog}. The Toda equation was used by Korepin and Zinn-Justin \cite{KZ} to 
derive the free energy of the six-vertex model with DWBC, assuming
some Ansatz on the behavior of subdominant terms in the large $N$
asymptotics of the free energy.

Another application of the Izergin determinantal formula is that
$\tau_N$ can be expressed in terms of  a partition function of
a random matrix model and also in terms of related orthogonal polynomials, see
 the paper  \cite{Z-J1} of Zinn-Justin.  In the ferroelectric phase the
expression in terms of orthogonal polynomials can be obtained as follows. 
For the evaluation of the Hankel determinant, let us
write $\phi(t)$ in the form of the Laplace transform of a discrete measure,
\begin{equation} \label{dph6}
\phi(t)=\frac{\sinh(2\ga)}{\sinh(t+\ga)\sinh(t-\ga)}=
4\sum_{l=1}^\infty e^{-2tl}\sinh(2\ga l).
\end{equation} 
Then
\begin{equation} \label{dph7}
\tau_n=\frac{2^{n^2}}{n!}\sum_{l_1,\ldots,l_n=1}^\infty \Delta(l_i)^2\prod_{i=1}^n
\left[2e^{-2tl_i}\sinh(2\ga l_i)\right], 
\end{equation}
where
\begin{equation} \label{dph8}
\Delta(l_i)=\prod_{1\le i<j\le n}(l_j-l_i)
\end{equation}
is the Vandermonde determinant. 

Introduce now discrete monic polynomials $P_j(x)=x^j+\dots$ orthogonal 
on the set
$\N=\{l=1,2,\ldots\}$ with respect to the weight,
\begin{equation} \label{dph9}
w(l)=2e^{-2tl}\sinh(2\ga l)=e^{-2tl+2\ga l}-e^{-2tl-2\ga l},
\end{equation}
so that
\begin{equation} \label{dph14}
\sum_{l=1}^\infty P_j(l)P_k(l)w(l)=h_k\delta_{jk}.
\end{equation}
Then it follows from (\ref{dph7}) that
\begin{equation} \label{dph15}
\tau_n=2^{n^2}\prod_{k=0}^{n-1}h_k, 
\end{equation}
see Appendix in the end of the paper. We will prove the following asymptotics of 
$h_k$.

\begin{theo} \label{hk}
For any $\ep>0$, as $k\to\infty$,
\begin{equation} \label{dph16}
h_k=\frac{(k!)^2q^{k+1}}{(1-q)^{2k+1}}\,\left(1+O(e^{-k^{1-\ep}})\right),
\end{equation}
where 
\begin{equation} \label{dph17}
q=e^{2\ga-2t}.
\end{equation}
The error term in (\ref{dph16}) is uniform on any compact subset of the set
\begin{equation} \label{dph18}
\left\{ (t,\ga):\; 0<\ga<t\right\}.
\end{equation}
\end{theo}

\subsection{Main result: Asymptotics of the partition function}

This work is a continuation of the work \cite{BF} of the first author with Vladimir Fokin.
In \cite{BF} the authors obtain the large $n$ asymptotics
of the partition function $Z_n$ in the disordered phase. They prove the conjecture 
of Paul Zinn-Justin \cite{Z-J1} that the large 
$n$ asymptotics of $Z_n$ in the disordered phase has the following form: for
some $\ep>0$,
\begin{equation}\label{main1}
Z_n=Cn^{\kappa}F^{n^2}[1+O(n^{-\ep})],
\end{equation}
and they find the exact value of the exponent $\kappa$,
\begin{equation}\label{main2}
\kappa=\frac{1}{12}-\frac{2\ga^2}{3\pi(\pi-2\ga)}\,.
\end{equation}
The value of $F$ in the disordered phase is given by the formula,
\begin{equation}\label{main3}
F=\frac{\pi[\sin(\ga+t)\sin(\ga-t)]}{2\ga\cos\frac{\pi t}{2\ga}}\,,
\end{equation}
the exact value of constant $C>0$ is not yet known.

Our main result
in the present paper is the following theorem.

\begin{theo} \label{main}
In the ferroelectric phase with $t>\ga>0$, for any $\ep>0$, as $n\to\infty$,
\begin{equation} \label{main4}
Z_n=C G^n F^{n^2}\left[1+O\left(e^{- n^{1-\ep}}\right)\right],
\end{equation}
where $C=1-e^{-4\ga}$, $G=e^{\ga-t}$, and $F=\sinh(t+\ga)$.
The error term in (\ref{dph16}) is uniform on any compact subset of the set
(\ref{dph18}).
\end{theo}

Up to a constant factor this result will follow from Theorem \ref{hk}. To find 
the constant factor $C$ we will use the Toda equation, combined with the asymptotics
of $C$ as $t\to\infty$. The proof of Theorems \ref{hk} and \ref{main} will be given below in 
Sections \ref{Meixner}-\ref{explicit_formula_C}. 
Here we would like to make some remarks concerning the phase transition
between the ferroelectric and disordered phases.

\subsection{Order of the phase transition between the ferroelectric and disordered phases}
We would like to compare the free energy in the disordered phase and in the
ferroelectric phase, when we approach a point of phase transition. Consider first
the ferroelectric phase. Observe that $t,\ga\to 0$ as we approach the line of phase
transition,
\begin{equation} \label{main5}
\frac{b}{c}=\frac{a}{c}+1,
\end{equation}
hence $a,b,c\to 0$ in parametrization (\ref{pf4}). Consider the regime, 
\begin{equation} \label{main6}
t,\ga\to +0,\qquad \frac{t}{\ga}\to \al>1.
\end{equation}
In this regime,
\begin{equation} \label{main7}
\lim_{\ga\to 0}\frac{b}{c}=\lim_{\ga\to 0}\frac{\sinh (t+\ga)}{\sinh(2\ga)}=\frac{\al+1}{2}\,,
\qquad \lim_{\ga\to 0}\frac{a}{c}=\lim_{\ga\to 0}\frac{\sinh (t-\ga)}{\sinh(2\ga)}=\frac{\al-1}{2}\,.
\end{equation}
We have to rescale formula (\ref{main4}) according to (\ref{cl_17}),
\begin{equation} \label{main8}
Z_n\left(\frac{a}{c},\frac{a}{c},\frac{b}{c},\frac{b}{c},1,1\right)
=c^{-n^2}Z_n(a,a,b,b,c,c)=C G^n F_0^{n^2}\left[1+O\left(e^{- n^{1-\ep}}\right)\right],
\end{equation}
where
\begin{equation} \label{main9}
F_0=\frac{F}{c}=\frac{\sinh(t+\ga)}{\sinh(2\ga)}\,.
\end{equation}
Similarly, in the disordered phase,
\begin{equation} \label{main10}
Z_n\left(\frac{a}{c},\frac{a}{c},\frac{b}{c},\frac{b}{c},1,1\right)
=C n^{\kappa} F_0^{n^2}[1+O(n^{-\ep})],
\end{equation}
where
\begin{equation} \label{main11}
F_0=\frac{F}{c}=\frac{\pi \sin(\ga-t)\sin(\ga+t)}{2\ga\sin(2\ga)\cos\frac{\pi t}{2\ga}}\,.
\end{equation}
Observe that parametrization (\ref{pf6}) in the disordered phase is not convenient
as we approach critical line (\ref{main5}). Namely, it corresponds to the limit when
\begin{equation} \label{main12}
t,\ga\to \frac{\pi}{2}-0,\qquad \frac{\frac{\pi}{2}-t}{\frac{\pi}{2}-\ga}\to \al>1.
\end{equation}
Therefore, we replace $t$ for $\frac{\pi}{2}-t$ and $\ga$ for $\frac{\pi}{2}-\ga$.
This gives the parametrization,
\begin{equation} \label{main13}
a=\sin(t-\ga),\quad b=\sin(t+\ga),\quad c=\sin(2|\ga|),\quad |\ga|<t.
\end{equation}
The approach to critical line (\ref{main5}) is described by regime (\ref{main6}).
Formula (\ref{main11}) reads in the new $t,\ga$ as 
\begin{equation} \label{main14}
F_0=\frac{\pi \sin(t-\ga)\sin(t+\ga)}{\left(\pi-2\ga\right)\sin(2\ga)
\cos\left[\frac{\pi(\frac{\pi}{2}-t)}{2(\frac{\pi}{2}-\ga)}\right]}\,.
\end{equation}
We consider $F_0$ on the line
\begin{equation} \label{main15}
\frac{a+b}{c}=\al,
\end{equation}
and we use the parameter
\begin{equation} \label{main17}
\be=\frac{b-a}{c}\,
\end{equation}
on this line.
In variables $\al,\be$,
\begin{equation} \label{main19}
F_0=\frac{\al+\be}{2}\;\;\textrm{in the ferroelectric phase},
\end{equation}
and
\begin{equation} \label{main20}
F_0=\frac{(\al+\be)g(t,\ga)}{2}\,\;\;\textrm{in the disordered phase},
\end{equation}
where
\begin{equation} \label{main21}
g(t,\ga)=\frac{\pi \sin(t-\ga)}{\left(\pi-2\ga\right)
\sin\left[\frac{\pi(t-\ga)}{(\pi-2\ga)}\right]}
\end{equation}
A straightforward calculation shows that
on the line $\frac{a+b}{c}=\al$ in the disordered phase, as $\be\to 1-0$,
\begin{equation} \label{main22}
g(t,\ga)=1+\frac{2(\al-1)^{3/2}(1-\be)^{3/2}}{3\pi(\al+1)^{1/2}}+O((1-\be)^2).
\end{equation}
By (\ref{main19}), $g(t,\ga)=1$ in the ferroelectric phase.
This implies that the free energy $F_0$ exhibits a phase
transition of the order $\frac{3}{2}$ with respect to the parameter $\be$ at the point 
$\be=1$. Fig.5
depicts the graph of $F_0=F_0(\be)$ (the left graph)
and its derivative, $F_0'(\be)$ (the right graph), as a function of $\be=\frac{b-a}{c}$ on the line 
$\frac{b+a}{c}=2$. 
Observe the square root singularities of $F_0'$ at $\be=\pm 1$, which correspond
to the phase transition of order $\frac{3}{2}\,.$
Since 
\begin{equation} \label{main18}
\Delta=\frac{a^2+b^2-c^2}{2ab}=\frac{\al^2+\be^2-2}{\al^2-\be^2}=1+\frac{4(\be-1)}{\al^2-1}+O((\be-1)^2)\,,
\end{equation}
it is a phase
transition of the order $\frac{3}{2}$ as well, with respect to the parameter $\De$ at the point 
$\De=1$.

\begin{center}
 \begin{figure}[h]\label{free_energy}
\begin{center}
   \scalebox{0.32}{\includegraphics[angle=-90]{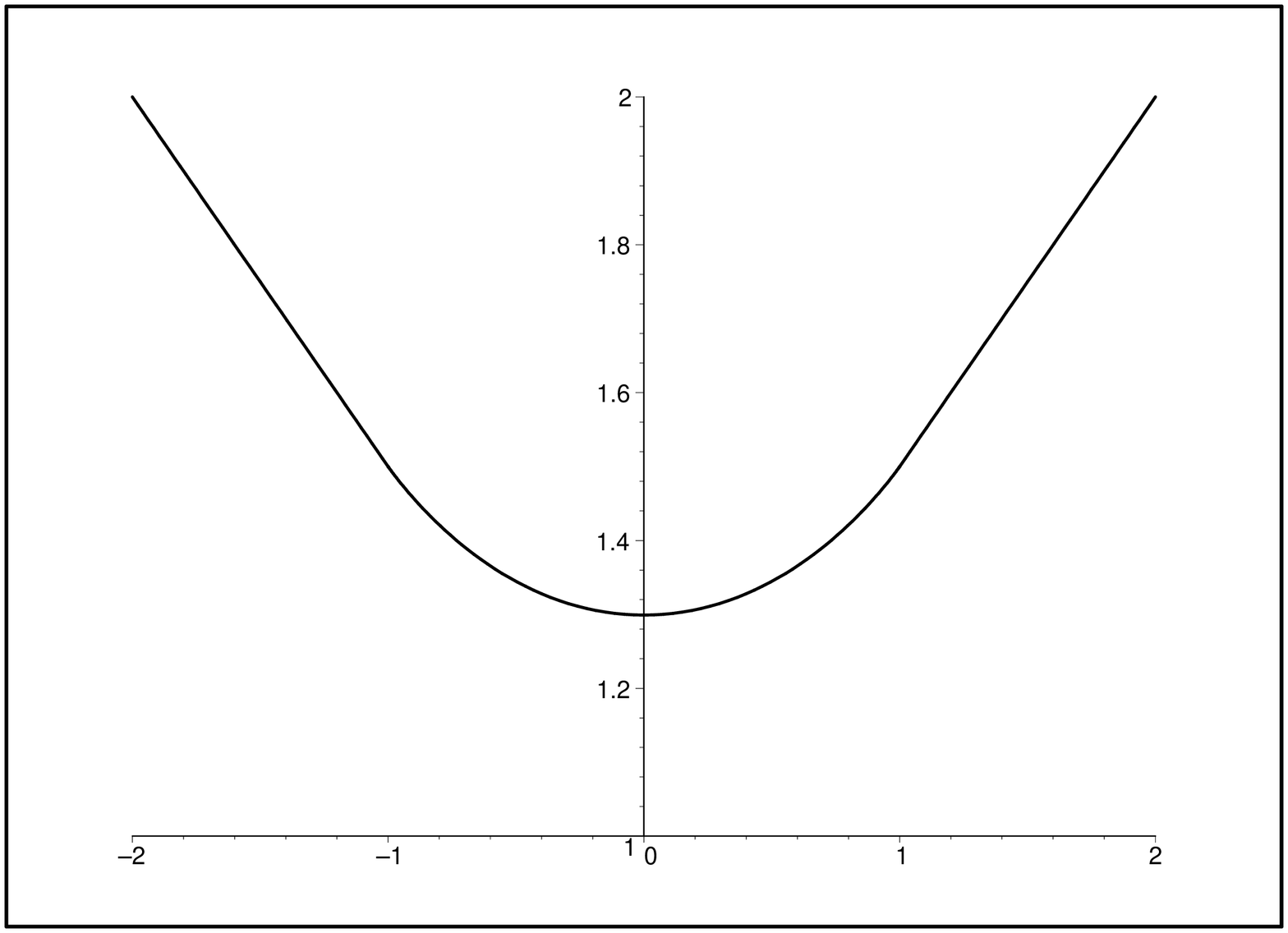}}
\scalebox{0.32}{\includegraphics[angle=-90]{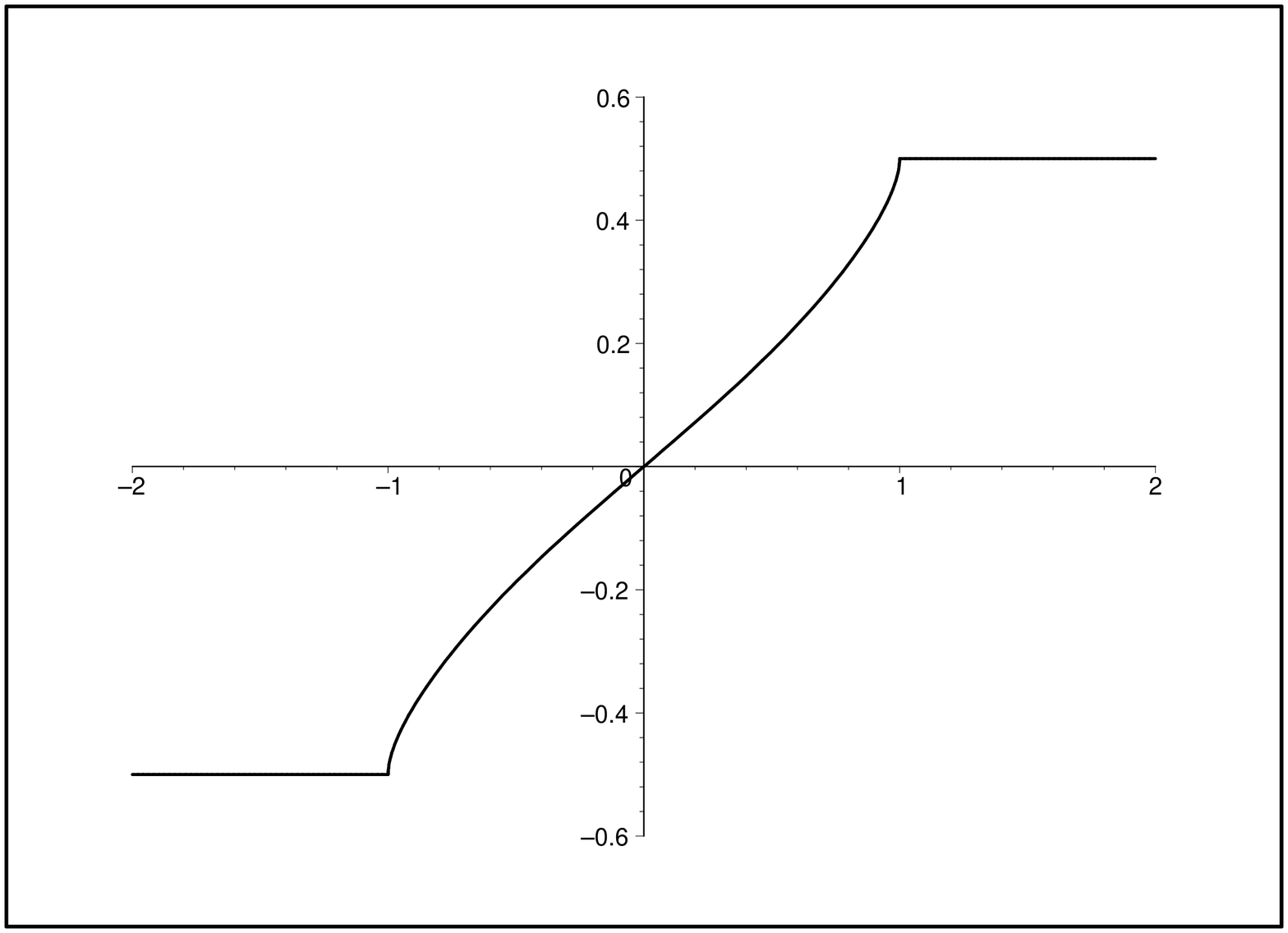}}
\end{center}
        \caption[Free energy ]
{Free energy $F_0=F_0(\be)$ (the left graph) and its derivative (the right graph), 
as functions of $\be=\frac{b-a}{c}$ on the line 
$\frac{b+a}{c}=2$. }
    \end{figure}
\end{center}

The set-up for the remainder of the article is the following. In Section \ref{Meixner}
we will discuss the Meixner polynomials, which will serve as a good approximation to the
polynomials $P_n(z)$. In Section \ref{RHA} we will discuss the Riemann-Hilbert approach 
to discrete orthogonal polynomials, and we will derive a basic identity, which will be used in the proof
of Theorem \ref{hk}. In Section \ref{hkk} we will prove Theorem \ref{hk}.  Then,
 in Sections \ref{evaluation_C}
and \ref{explicit_formula_C} we will obtain an explicit formula for the constant factor $C$,
and we will finish the proof of Theorem \ref{main}.
Finally, in Section
\ref{ground_state} we will compare the asymptotics of the free energy in the
ferroelectric phase with the energy of the ground state configuration.

\section{Meixner polynomials}\label{Meixner}

We will use the two weights: the weight $w(l)$ defined in (\ref{dph9})
and the exponential weight on $\N$,
\begin{equation} \label{meix0}
w^{\rm Q}(l)=q^l,\quad l\in\N;\qquad q=e^{2\ga-2t}<1,
\end{equation}
which can be viewed as an approximation to $w(l)$ for large $l$.
The orthogonal polynomials with the weight $w^{\rm Q}(l)$ are expressed 
in terms of the Meixner polynomials with $\be=1$, which are defined by
the formula,
\begin{equation} \label{meix1}
\begin{aligned}
M_k(z;q)&= {}_2F_1\left(\begin{matrix} -k,-z \\ 1 \end{matrix};1-q^{-1}\right)
=\sum_{j=0}^{\infty}\frac{(-k)_j(-z)_j}{(1)_j}\,\frac{(1-q^{-1})^j}{j!}\\
&=\sum_{j=0}^{k}\frac{(1-q^{-1})^j \prod_{i=0}^{j-1}(k-i) \prod_{i=0}^{j-1}(z-i)}{ (j!)^2}\,.
\end{aligned}
\end{equation}
They satisfy the orthogonality condition,
\begin{equation} \label{meix2}
\sum_{l=0}^\infty M_j(l;q)M_k(l;q)q^l= \frac{q^{-k}\de_{jk}}{1-q}\,,
\end{equation}
see, e.g. \cite{KS}. For the corresponding monic polynomials,
\begin{equation} \label{meix3}
P_k^{\rm M}(z)=\frac{k!}{(1-q^{-1})^k} M_k(z;q)
\end{equation}
(M in $P_k^{\rm M}$ stands for Meixner), the orthogonality condition reads 
\begin{equation} \label{meix4}
\sum_{l=0}^\infty P_j^{\rm M}(l)P_k^{\rm M}(l)q^j=h_k^{\rm M}\de_{jk},
\qquad h_k^{\rm M}= \frac{(k!)^2q^{k}}{(1-q)^{2k+1}}\,.
\end{equation} 
They satisfy the three term recurrence relation,
\begin{equation} \label{meix5}
zP_k^{\rm M}(z)=P_{k+1}^{\rm M}(z)+\frac{kq+k+q}{1-q}\,P_k^{\rm M}(z)
+\frac{k^2q}{(1-q)^2}\,P_{k-1}^{\rm M}(z),
\end{equation}
see \cite{KS}. According to (\ref{IP7}), we take
$q=e^{2\ga-2t}$.

For our purposes it is convenient to introduce a shifted Meixner polynomial,
\begin{equation} \label{meix6}
Q_k(z)=P_k^{\rm M}(z-1)=\frac{(-1)^k k!q^k}{(1-q)^k}M_k(z-1;q),
\end{equation}
which is a monic polynomial as well.
Equation (\ref{meix4}) implies the orthogonality condition,
\begin{equation} \label{meix7}
\sum_{l=1}^\infty Q_j(l)Q_k(l)q^l=h_k^{\rm Q}\de_{jk},
\qquad h_k^{\rm Q}= \frac{(k!)^2q^{k+1}}{(1-q)^{2k+1}}\,.
\end{equation} 
By analogy with (\ref{dph15}), define
\begin{equation} \label{meix8}
\tau_n^{\rm Q}=2^{n^2}\prod_{k=0}^{n-1}h_k^{\rm Q}.
\end{equation}
From (\ref{meix4}) and (\ref{meix7}) we obtain that
\begin{equation} \label{meix9}
\tau_n^{\rm Q}=2^{n^2}\prod_{k=0}^{n-1}\frac{(k!)^2q^{k+1}}{(1-q)^{2k+1}}
=\frac{2^{n^2} q^{(n+1)n/2}}{(1-q)^{n^2}}\prod_{k=0}^{n-1}(k!)^2\,. 
\end{equation}
By analogy with (\ref{pf7}), define also
\begin{equation} \label{meix10}
Z_n^{\rm Q}=\frac{[\sinh(\ga+t)\sinh(\ga-t)]^{n^2}}{
\displaystyle \prod_{k=0}^{n-1}(k!)^2}\,\tau_n^{\rm Q}\,.
\end{equation}
Then from (\ref{meix9}) we obtain that
\begin{equation} \label{meix11}
Z_n^{\rm Q}=F^{n^2}G^n,
\end{equation}
where
\begin{equation} \label{meix12}
F=\frac{2\sinh(t-\ga)\sinh(t+\ga)q^{1/2}}{1-q}
=\frac{2\sinh(t-\ga)\sinh(t+\ga)e^{\ga-t}}{1-e^{2\ga-2t}}=\sinh(t+\ga),
\end{equation}
and 
\begin{equation} \label{meix13}
G=q^{1/2}=e^{\ga-t}.
\end{equation}
Our goal will be to compare the normalizing constants for
orthogonal polynomials with the weights $w$ and $w^{\rm Q}$.
To this end let us discuss the Riemann-Hilbert approach to discrete
orthogonal polynomials. 

\section{Riemann Hilbert approach: Interpolation problem}\label{RHA}

The Riemann-Hilbert approach to discrete orthogonal polynomials
 is based on the following Interpolation Problem (IP), which was
introduced in the paper \cite{BoB} of Borodin and Boyarchenko
under the name of the discrete Riemann-Hilbert problem.
See also the monograph  \cite{BKMM}
of Baik, Kriecherbauer, McLaughlin, and Miller, in which it is called the
Interpolation Problem.  
Let $w(l)\ge 0$ be a weight function on $\N$ (it can be a more general
discrete set, as discussed in \cite{BoB} and \cite {BKMM}, but we will
need $\N$ in our problem).

{\bf Interpolation Problem}. For a given $k=0,1,\ldots$, find a $2\times 2$ matrix-valued function
$Y(z;k)=(Y_{ij}(z;k))_{1\le i,j\le 2}$ with the following properties:
\begin{enumerate}
\item
{\bf Analyticity}: $Y(z;k)$ is an analytic function of $z$ for $z\in\C\setminus\N$.
\item
{\bf Residues at poles}: At each node $l\in\N$, the elements $Y_{11}(z;k)$ and
$Y_{21}(z;k)$ of the matrix $Y(z;k)$ are analytic functions of $z$, and the elements $Y_{12}(z;k)$ and
$Y_{22}(z;k)$ have a simple pole with the residues,
\begin{equation} \label{IP1}
\underset{z=l}{\rm Res}\; Y_{j2}(z;k)=w(l)Y_{j1}(l;k),\quad j=1,2.
\end{equation}
\item
{\bf Asymptotics at infinity}: As $z\to\infty$, $Y(z;k)$ admits the asymptotic expansion,
\begin{equation} \label{IP2}
Y(z;k)\sim \left( I+\frac {Y_1}{z}+\frac {Y_2}{z^2}+\ldots\right)
\begin{pmatrix}
z^k & 0 \\
0 & z^{-k}
\end{pmatrix},\qquad z\in \C\setminus \left[\bigcup_{l=1}^\infty D(l,r_l)\right],
\end{equation}
where $D(z,r)$ is a disk of radius $r>0$ centered at $z\in \C$ and 
\begin{equation} \label{IP2a}
\lim_{l\to\infty} r_l=0.
\end{equation}
\end{enumerate}

It is not difficult to see (see \cite{BoB} and \cite{BKMM}) that under some conditions on $w(l)$,
the IP has a unique solution, which is
\begin{equation} \label{IP3}
Y(z;k)=
\begin{pmatrix}
P_k(z) & C(wP_k)(z) \\
(h_{k-1})^{-1}P_{k-1}(z) & (h_{k-1})^{-1}C(wP_{k-1})(z)
\end{pmatrix}
\end{equation}
where the Cauchy transformation $C$ is defined by the formula,
\begin{equation} \label{IP4}
C(f)(z)=\sum_{l=1}^\infty\frac{f(l)}{z-l}\,,
\end{equation}
and $P_k(z)=z^k+\ldots$ are monic polynomials orthogonal with the weight $w(l)$,
so that
\begin{equation} \label{IP5}
\sum_{l=1}^\infty P_j(l)P_k(l)w(l)=h_j\delta_{jk}.
\end{equation}
It follows from (\ref{IP3}), that 
\begin{equation} \label{IP6}
h_k=[Y_1]_{21},
\end{equation}
where $[Y_1]_{21}$ is the (21)-element of the matrix $Y_1$ on the right in (\ref{IP2}).
In what follows we will consider the solution
$Y(z;k)$ for the weight $w(l)$, introduced in (\ref{dph9}). 

Let $Y^{\rm Q}$ be  a solution to the IP with the exponential weight $w^{\rm Q}$,
\begin{equation} \label{IP7}
Y^{\rm Q}(z;k)=
\begin{pmatrix}
Q_k(z) & C(w^{\rm Q}Q_k)(z) \\
(h_{k-1}^{\rm Q})^{-1}Q_{k-1}(z) & (h_{k-1}^{\rm Q})^{-1}C(w^{\rm Q}Q_{k-1})(z)
\end{pmatrix}.
\end{equation}
Consider the quotient matrix,
\begin{equation} \label{IP8}
X(z;k)=Y(z;k)[Y^{\rm Q}(z;k)]^{-1}.
\end{equation}
Observe that $\det Y^{\rm Q}(z;k)$ has no poles and it approaches 1 as $z\to \infty $
outside of the disks $D(l,r_l)$, $l=1,2,\ldots$, hence
\begin{equation} \label{IP9}
\det Y^{\rm Q}(z;k)=1.
\end{equation}
Also, 
\begin{equation} \label{IP10}
X(z;k)\to I\quad \textrm{as $z\to \infty $
outside of the disks $D(l,r_l)$, $l=1,2,\ldots$}
\end{equation}
This implies that the matrix $X$ can be written as
\begin{equation} \label{IP11}
X(z;k)=I+C[(w^{\rm Q}-w)R],
\end{equation}
where
\begin{equation} \label{IP12}
R(z)=
\begin{pmatrix}
(h_{k-1}^{\rm Q})^{-1}P_k(z)Q_{k-1}(z) & -P_k(z)Q_k(z) \\
(h_{k-1}h_{k-1}^{\rm Q})^{-1}P_{k-1}(z)Q_{k-1}(z) & -(h_{k-1})^{-1}P_{k-1}(z)Q_k(z)
\end{pmatrix}.
\end{equation}
From formula (\ref{IP6}) and (\ref{IP11}) we obtain that
\begin{equation} \label{IP13}
h_k-h_k^{\rm Q}=-\sum_{l=1}^\infty P_k(l)Q_k(l)\,[w^{\rm Q}(l)-w(l)].
\end{equation}
We will use this identity to estimate $|h_k-h_k^{\rm Q}|$.
Observe that formula (\ref{IP11}) can be further used to evaluate the large
$n$ asymptotics of the orthogonal polynomials $P_n(z)$, but we will not pursue
it here.

We would like to remark that identity (\ref{IP13}) can be also derived as follows.
Observe that since $P_k$ and $Q_k$ are monic polynomials, the difference, $P_k-Q_k$,
is a polynomial of degree less than $k$, hence
\begin{equation} \label{IP14}
\sum_{l=1}^\infty P_k(l)[Q_k(l)-P_k(l)]w(l)=0.
\end{equation}
By adding this to equation (\ref{IP5}) with $j=k$, we obtain that
\begin{equation} \label{IP15}
h_k=\sum_{l=1}^\infty P_k(l)Q_k(l)w(l).
\end{equation}
Similarly, from (\ref{meix7}) we obtain that
\begin{equation} \label{IP16}
h_k^{\rm Q}=\sum_{l=1}^\infty P_k(l)Q_k(l)w^{\rm Q}(l).
\end{equation} 
By subtracting the last two equations, we obtain identity (\ref{IP13}).

\section{Evaluation of the ratio $h_k/h_k^{\rm Q}$}\label{hkk}

In this section we will prove Theorem \ref{hk}. 
By applying the Cauchy-Schwarz inequality to identity (\ref{IP13}), we obtain that
\begin{equation} \label{h5}
|h_k-h_k^{\rm Q}|\le\left[\sum_{l=1}^\infty P_k(l)^2\,|w(l)-w^{\rm Q}(l)|\right]^{1/2}
\left[\sum_{l=1}^\infty Q_k(l)^2\,|w(l)-w^{\rm Q}(l)|\right]^{1/2},
\end{equation}
so that
\begin{equation} \label{h6}
\left|\frac{h_k}{h_k^{\rm Q}}-1\right|\le\left[\frac{1}{h_k^{\rm Q}}
\sum_{l=1}^\infty P_k(l)^2\,|w(l)-w^{\rm Q}(l)|\right]^{1/2}
\left[\frac{1}{h_k^{\rm Q}}\sum_{l=1}^\infty Q_k(l)^2\,|w(l)-w^{\rm Q}(l)|\right]^{1/2},
\end{equation}
From (\ref{dph9}),
\begin{equation} \label{h8}
|w(l)-w^{\rm Q}(l)|=e^{-(2t+2\ga)l}\le C_0w(l),\quad l\ge 1;\qquad C_0=\frac{1}{e^{4\ga}-1}\,,
\end{equation}
hence
\begin{equation} \label{h9}
\frac{1}{h_k^{\rm Q}}
\sum_{l=1}^\infty P_k(l)^2\,|w(l)-w^{\rm Q}(l)|
\le C_0\frac{1}{h_k^{\rm Q}}
\sum_{l=1}^\infty P_k(l)^2 w(l)=\frac{C_0 h_k}{h_k^{\rm Q}}\le C_0(1+\ep_k),
\end{equation}
where
\begin{equation} \label{h10}
\ep_k=\left|\frac{h_k}{h_k^{\rm Q}}-1\right|\,.
\end{equation}
Thus, by (\ref{h6}),
\begin{equation} \label{h11}
\ep_k^2\le C_0(1+\ep_k)\de_k,
\end{equation}
where
\begin{equation} \label{h12}
\de_k=\frac{1}{h_k^{\rm Q}}\sum_{l=1}^\infty Q_k(l)^2\,|w(l)-w^{\rm Q}(l)|
\end{equation}
By (\ref{h8}),
\begin{equation} \label{h13}
\de_k=\frac{1}{h_k^{\rm Q}}\sum_{l=1}^\infty Q_k(l)^2\,q_0^l,
\qquad q_0=e^{-2(t+\ga)}.
\end{equation}
Let us evaluate $\de_k$.

We partition the sum in (\ref{h13}) into two parts:
\begin{equation} \label{h14}
\de_k'=\frac{1}{h_k^{\rm Q}}\sum_{l=1}^L Q_k(l)^2\,q_0^l,
\end{equation}
and
\begin{equation} \label{h15}
\de_k''=\frac{1}{h_k^{\rm Q}}\sum_{l=L+1}^\infty Q_k(l)^2\,q_0^l,
\end{equation}
where 
\begin{equation} \label{h15a}
L=[k^\la],\qquad 0<\la<1.
\end{equation}
Let us estimate first $\de_k'$. We have from (\ref{meix6}), (\ref{meix7}) that
\begin{equation} \label{h16}
\frac{Q_k(l)}{(h_k^{\rm Q})^{1/2}}=\frac{(-1)^k(1-q)^{1/2}q^{k/2}}{q^{1/2}}M_k(l-1;q).
\end{equation}
By (\ref{meix1}),
\begin{equation} \label{h17}
\begin{aligned}
M_k(l-1;q)&=1+(1-q^{-1})k(l-1)+(1-q^{-1})^2\frac{k(k-1)(l-1)(l-2)}{(2!)^2}\\
&+(1-q^{-1})^3\frac{k(k-1)(k-2)(l-1)(l-2)(l-3)}{(3!)^2}+\ldots
\end{aligned}
\end{equation}
If $l<k$, then the latter sum consists of $l$ nonzero terms. For $l\le L$ it is estimated as 
\begin{equation} \label{h18}
M_k(l-1;q)=O(k^L L^{L+1})=O(e^{L\ln k+(L+1)\ln L}),
\end{equation}
hence
\begin{equation} \label{h19}
\frac{Q_k(l)}{(h_k^{\rm Q})^{1/2}}=O(e^{\frac{k\ln q}{2}+L\ln k+(L+1)\ln L}).
\end{equation}
Due to our choice of $L$ in (\ref{h15a}), this implies the estimate,
\begin{equation} \label{h19a}
\frac{Q_k(l)}{(h_k^{\rm Q})^{1/2}}=O(e^{\frac{k\ln q}{2}+2k^\la \ln k}).
\end{equation}
Since $0<q<1$ and $0<\la<1$, the expression on the right is exponentially
small as $k\to\infty$.
From (\ref{h14}) we obtain  now that
\begin{equation} \label{h20}
\de_k'=O(e^{k\ln q+4k^\la \ln k}).
\end{equation}
Since $\la<1$ and $q<1$, we obtain that
\begin{equation} \label{h22}
\de_k'=O(e^{-c_0 k}),\qquad c_0=-\frac{\ln q}{2}>0.
\end{equation}
Let us estimate $\de_k''$.

By (\ref{meix7}),
\begin{equation} \label{h23}
\frac{1}{h_k^{\rm Q}}\sum_{l=1}^\infty Q_k(l)^2q^l=1,
\end{equation} 
hence
\begin{equation} \label{h24}
\de_k''=\frac{1}{h_k^{\rm Q}}\sum_{l=L+1}^\infty Q_k(l)^2\,q_0^l
< \left(\frac{q_0}{q}\right)^L\frac{1}{h_k^{\rm Q}}\sum_{l=L+1}^\infty Q_k(l)^2\,q^l
<\left(\frac{q_0}{q}\right)^L=e^{-4\ga L}.
\end{equation}
Thus,
\begin{equation} \label{h25}
\de_k''<e^{-4\ga (k^\la-1)}.
\end{equation}
Since $0<\la<1$ is an arbitrary number, we obtain from (\ref{h22}) and (\ref{h25}) that
for any $\eta>0$,
\begin{equation} \label{h26}
\de_k=O\left(e^{- k^{1-\eta}}\right).
\end{equation}
Let us return back to inequality (\ref{h11}). Consider two cases: (1) $\ep_k>1$ and (2) $\ep_k\le 1$.
In the first case (\ref{h11}) implies that
\begin{equation} \label{h27}
\ep_k\le 2C_0\de_k,
\end{equation}
which is impossible, because of (\ref{h26}). Hence $\ep_k\le 1$, in which case (\ref{h11}) gives
that
\begin{equation} \label{h28}
\ep_k^2\le 2C_0\de_k.
\end{equation}
Estimate (\ref{h26}) implies now that for any $\eta>0$,
\begin{equation} \label{h29}
\ep_k=O\left(e^{- k^{1-\eta}}\right),
\end{equation}
so that as $k\to\infty$,
\begin{equation} \label{h30}
h_k=h_k^{\rm Q}(1+\tilde\ep_k),\qquad |\tilde\ep_k|=\ep_k=O\left(e^{- k^{1-\eta}}\right).
\end{equation}
This proves Theorem \ref{hk}.

From (\ref{h30}) we obtain that for any $\eta>0$,
\begin{equation} \label{h31}
Z_n=Z_n^{\rm Q}\prod_{k=0}^n(1+\tilde\ep_k)
=C Z_n^{\rm Q}\left[1+O\left(e^{- n^{1-\eta}}\right)\right],
\end{equation}
where
\begin{equation} \label{h32}
C=\prod_{k=0}^\infty(1+\tilde\ep_k)>0.
\end{equation}
Thus, we have proved the following result.

\begin{prop} For any $\ep>0$, as $n\to\infty$,
\begin{equation} \label{h33}
Z_n=C F^{n^2}G^n\left[1+O\left(e^{- n^{1-\ep}}\right)\right],
\end{equation}
where $C>0$, $F=\sinh(t+\ga)$, and $G=e^{\ga-t}$.
\end{prop}

To finish the proof of Theorem \ref{main}, it remains to find the constant $C$.

\vskip 3mm

\section{Evaluation of the constant factor}\label{evaluation_C}

In the next two sections we will find the exact value of the constant $C$ in 
formula (\ref{h33}). This will be done in two steps: first, with the help
of the Toda equation, we will find the form of the dependence of $C$ on $t$,
and second, we will find the large $t$ asymptotics of $C$. By combining 
these two steps, we will obtain the exact value of $C$. In this section
we will carry out the first step of our program.

By dividing the Toda equation, (\ref{pf10}), by $\tau_n^2$, we obtain that
\begin{equation} \label{co1}
\frac{\tau_n\tau_n''-\tau_n'^2}{\tau_n^2}=\frac{\tau_{n+1}\tau_{n-1}}{\tau_n^2}\,,
\qquad (')=\frac{\partial}{\partial t}\,.
\end{equation}
The left hand side can be written as
\begin{equation} \label{co2}
\frac{\tau_n\tau_n''-\tau_n'^2}{\tau_n^2}=\left(\frac{\tau_n'}{\tau_n}\right)'
=\left(\ln\tau_n\right)''.
\end{equation}
From (\ref{dph15}) we obtain that
\begin{equation} \label{co3}
\frac{\tau_{n+1}}{\tau_n}=2^{2n+1}h_n,
\end{equation}
hence equation (\ref{co1}) implies that
\begin{equation} \label{co4}
\left(\ln\tau_n\right)''=\frac{4h_n}{h_{n-1}}\,.
\end{equation}
From (\ref{dph16}) we obtain that
\begin{equation} \label{co5}
\frac{4h_n}{h_{n-1}}=\frac{4n^2q}{(1-q)^2}+O\left(e^{-n^{1-\ep}}\right).
\end{equation}
We have that
\begin{equation} \label{co6}
\frac{4q}{(1-q)^2}=\frac{4e^{2\ga-2t}}{(1-e^{2\ga-2t})^2}=\left[\frac{(-2)}{1-e^{2\ga-2t}}\right]'
=\left[-\ln(1-e^{2\ga-2t})\right]'',
\end{equation}
hence from (\ref{co4}), (\ref{co5}) we obtain that 
\begin{equation} \label{co7}
\left(\ln\tau_n\right)''=n^2\left[-\ln(1-e^{2\ga-2t})\right]''+O\left(e^{-n^{1-\ep}}\right).
\end{equation}
By (\ref{pf7}) this implies that
\begin{equation} \label{co8}
\left(\ln Z_n\right)''=n^2\left[\ln\frac{\sinh(t-\ga)\sinh(t+\ga)}{1-e^{2\ga-2t}}\right]''
+O\left(e^{-n^{1-\ep}}\right).
\end{equation}
Since
\begin{equation} \label{co9}
\ln\frac{\sinh(t-\ga)\sinh(t+\ga)}{1-e^{2\ga-2t}}=\ln[\sinh(t+\ga)]+(t-\ga)-\ln 2,
\end{equation}
we can simplify (\ref{co8}) to
\begin{equation} \label{co10}
\left(\ln Z_n\right)''=n^2\left[\ln\sinh(t+\ga)\right]''
+O\left(e^{-n^{1-\ep}}\right).
\end{equation}
Observe that the error term in the last formula
is uniform when $t$ belongs to a compact set on $(\ga,\infty)$,
hence by integrating it we obtain that 
\begin{equation} \label{co11}
\ln Z_n=n^2\ln\sinh(t+\ga)+c_1t+c_0
+O\left(e^{-n^{1-\ep}}\right),
\end{equation}
where $c_0,c_1$ do not depend on $t$. In general, $c_0,c_1$ depend on $\ga$ and $n$.
By substituting formula (\ref{h33}) into the preceding equation, we obtain that
\begin{equation} \label{co12}
\ln C+n(\ga-t)=c_1t+c_0
+O\left(e^{-n^{1-\ep}}\right).
\end{equation}
Denote 
\begin{equation} \label{co13}
d_0=c_0-n\ga,\qquad d_1=c_1+n.
\end{equation}
Then  equation (\ref{co12}) reads
\begin{equation} \label{co14}
\ln C=d_1t+d_0
+O\left(e^{-n^{1-\ep}}\right).
\end{equation}
Observe that $C=C(\ga,t)$ does not depend on $n$, while
$d_j=d_j(\ga,n)$ does not depend on $t$, $j=1,2$.  Take any $0<\ga<t_1<t_2$. Then
\begin{equation} \label{co15}
\ln C(\ga,t_2)-\ln C(\ga,t_1)=d_1(t_2-t_1)
+O\left(e^{-n^{1-\ep}}\right).
\end{equation}
From this formula we obtain that the limit,
\begin{equation} \label{co16}
\lim_{n\to\infty} d_1(\ga,n)=d_1(\ga),
\end{equation}
exists. This in turn implies that the limit,
\begin{equation} \label{co17}
\lim_{n\to\infty} d_2(\ga,n)=d_2(\ga),
\end{equation}
exists. By taking the limit $n\to\infty$ in (\ref{co14}), we obtain that
\begin{equation} \label{co18}
\ln C=d_1(\ga)t+d_0(\ga).
\end{equation}
Thus we have proved the following result.

\begin{prop} The constant $C$ in asymptotic formula (\ref{h33}) has the form
\begin{equation} \label{co19}
C=e^{d_1(\ga)t+d_0(\ga)}.
\end{equation}
\end{prop}

\section{Explicit formula for $C$} \label{explicit_formula_C}

In this section we will find the exact value of $C$, and by doing this we will
finish the proof of Theorem \ref{main}. 
Let us consider the following regime:
\begin{equation} \label{forC1}
\ga\; \textrm{is fixed},\quad t\to\infty,
\end{equation}
and let us evaluate the asymptotics of $C$ in this regime.
By (\ref{IP5}) and (\ref{dph9}) we have that
\begin{equation} \label{forC2}
h_0=\sum_{l=1}^\infty w(l)=\sum_{l=1}^\infty \left(e^{-2tl+2\ga l}-e^{-2tl-2\ga l}\right)
=\frac{e^{-2t+2\ga }}{1-e^{-2t+2\ga }}-\frac{e^{-2t-2\ga }}{1-e^{-2t-2\ga }}\,.
\end{equation}
Similarly, by (\ref{meix7}),
\begin{equation} \label{forC3}
h_0^{\rm Q}
=\frac{e^{-2t+2\ga }}{1-e^{-2t+2\ga }}\,,
\end{equation}
hence
\begin{equation} \label{forC4}
\frac{h_0}{h_0^{\rm Q}}=1-e^{-4\ga}+O(e^{-2t}),\qquad t\to\infty.
\end{equation}
Let us evaluate $\ep_k=\left|\frac{h_k}{h_k^{\rm Q}}-1\right|$ for $k\ge 1$.

By (\ref{h11}),
\begin{equation} \label{forC5}
\ep_k^2\le C_0(1+\ep_k)\de_k,\qquad C_0=\frac{1}{e^{4\ga}-1}\,.
\end{equation}
In the partition of $\de_k$ as $\de_k'+\de_k''$ in (\ref{h14}), (\ref{h15}),
let us choose
\begin{equation} \label{forC6}
L=[k^{2/3}+t^{2/3}].
\end{equation}
From (\ref{h16}), (\ref{h17}) we obtain that for $l\le L$,
\begin{equation} \label{forC7}
\frac{|Q_k(l)|}{(h_k^{\rm Q})^{1/2}}\le  q^{(k-1)/2}k^L L^{L+1},\qquad q=e^{2\ga-2t},
\end{equation}
hence
\begin{equation} \label{forC8}
\de_k'\le  \frac{q_0q^{k-1}k^L L^{L+1}}{1-q_0}\le  \frac{q^{k}k^L L^{L+1}}{1-q_0}\,,
\qquad q_0=e^{-2\ga-2t}.
\end{equation}
In addition, by (\ref{h24}),
\begin{equation} \label{forC9}
\de_k''\le  e^{-4\ga L}.
\end{equation}
Our choice of $L$ in (\ref{forC6}) ensures that there exists $t_0>0$
such that for any $t\ge t_0$ and any $k\ge 1$,
\begin{equation} \label{forC10}
\de_k=\de_k'+\de_k''\le  e^{-k^{1/2}-t^{1/2}}.
\end{equation}
From (\ref{forC5}) we obtain now that for $k\ge 1$ and large $t$,
\begin{equation} \label{forC11}
\ep_k\le C_1 e^{-\frac{k^{1/2}}{2}-\frac{t^{1/2}}{2}},\qquad C_1=(2C_0)^{1/2}.
\end{equation}
By (\ref{h32}),
\begin{equation} \label{forC12}
\ln C=\sum_{k=0}^\infty \ln(1+\tilde\ep_k),\qquad |\tilde\ep_k|=\ep_k.
\end{equation}
From equations (\ref{forC4}) and (\ref{forC11}) we obtain now that 
\begin{equation} \label{forC13}
\ln C=\ln(1-e^{-4\ga})+O(e^{-\frac{t^{1/2}}{2}}),\qquad t\to\infty.
\end{equation} 
On the other hand, by (\ref{co14})
\begin{equation} \label{forC14}
\ln C=d_1(\ga)t+d_0(\ga)
\end{equation} 
This implies that
\begin{equation} \label{forC15}
d_1(\ga)=0,\qquad d_0(\ga)=\ln(1-e^{-4\ga}),
\end{equation} 
so that
\begin{equation} \label{forC16}
C=1-e^{-4\ga}.
\end{equation} 
By substituting expression (\ref{forC16}) into formula (\ref{h33}), we prove 
Theorem \ref{main}.

Let us compare now the asymptotics of the free energy in the ferroelectric phase with the
energy of the ground state.
 
\section{Ground state configuration of the ferroelectric phase}\label{ground_state}

\begin{center}
 \begin{figure}[h]\label{gs}
\begin{center}
   \scalebox{0.7}{\includegraphics{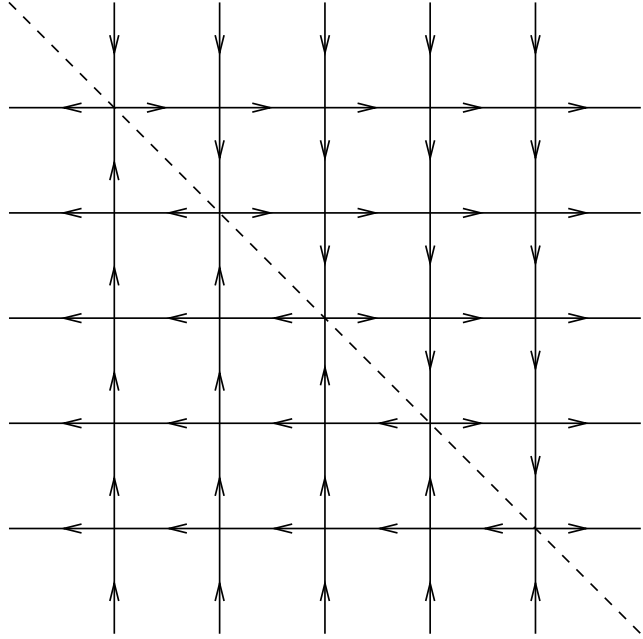}}
\end{center}
        \caption[A ground state configuration. ]
{A ground state configuration. }
    \end{figure}
\end{center}
The ground state is the configuration
\begin{equation} \label{gs1}
\sg^{\rm gs}(x)=\left\{
\begin{aligned}
&\sg_5\quad{\rm if }\; x\;\text{\rm is on the diagonal},\\
&\sg_3\quad{\rm if }\; x\;\text{\rm is above the diagonal},\\
&\sg_4\quad{\rm if }\; x\;\text{\rm is below the diagonal},
\end{aligned}\right.
\end{equation}
see Fig.5. The weight of the ground state configuration is
\begin{equation} \label{gs2}
w(\sg^{\rm gs})=b^{n^2}\left(\frac{c}{b}\right)^n=F^{n^2}G_0^n,
\end{equation}
where
\begin{equation} \label{gs3}
F=\sinh(t+\ga),\qquad G_0=\frac{\sinh(2\ga)}{\sinh(t+\ga)}.
\end{equation}
The ratio $Z_n/w(\sg^{\rm gs})$ is evaluated as
\begin{equation} \label{gs4}
\frac{Z_n}{w(\sg^{\rm gs})}=G_1^n,
\end{equation}
where
\begin{equation} \label{gs5}
G_1=\frac{G}{G_0}=\frac{e^{\ga-t}\sinh(t+\ga)}{\sinh{2\ga}}=\frac{e^{2\ga}-e^{-2t}}{e^{2\ga}-e^{-2\ga}}>1.
\end{equation}
Observe that 
\begin{equation} \label{gs6}
\lim_{n\to\infty}\frac{\ln Z_n}{n^2}=\lim_{n\to\infty}\frac{\ln w(\sg^{\rm gs})}{n^2}=\ln F,
\end{equation}
so that the free energy is determined by the free energy
of the ground state configuration. This can be explained by the fact that
low energy excited states are local perturbations of the ground state around the diagonal. Namely, it
is impossible to create a new configuration by perturbing the ground state locally away of the diagonal:
the conservation law $N_3(\sg)=N_4(\sg)$ forbids such a configuration. Therefore, typical configurations
of the six-vertex model in the ferroelectric phase are frozen outside of a relatively
small neighborhood of the diagonal. 

This behavior of typical configurations in
the ferroelectric phase is in a big contrast with
the situation in the disordered and anti-ferroelectric phases. Extensive rigorous, theoretical
and numerical studies, see, e.g., the works of  Cohn, Elkies, Propp \cite{CEP},
Eloranta \cite{Elo}, 
Syljuasen,  Zvonarev  \cite{SZ}, Allison, Reshetikhin \cite{AR}, Kenyon, Okounkov \cite{KO2},
Kenyon, Okounkov, Sheffield \cite{KOS}, Sheffield \cite{She},
Ferrari, Spohn \cite{FS}, Colomo, Pronko \cite{CP3},
Zinn-Justin \cite{Z-J2},
and references therein, show that in the disordered and anti-ferroelectric phases the
``arctic circle'' phenomenon persists, so that there are macroscopically big
frozen and random domains    in typical configurations, separated in the limit $n\to\infty$
by an ``arctic curve''.
\medskip

\appendix

\section{Derivation of formula (\ref{dph15})}

Multilinearity of the determinant function, combined with the form of the Vandermonde matrix, allows us 
to replace $\Delta(l_{i})$ with  
\begin{equation} \label{app1}
\det\begin{pmatrix}
1 & 1 & 1 & \cdots & 1 \\
P_{1}(l_{1}) & P_{1}(l_{2}) & P_{1}(l_{3}) & \cdots & P_{1}(l_{n}) \\
P_{2}(l_{1}) & P_{2}(l_{2}) & P_{2}(l_{3}) & \cdots & P_{2}(l_{n}) \\
\vdots&\vdots&\vdots&\vdots&\vdots \\
P_{n-1}(l_{1}) & P_{n-1}(l_{2}) & P_{n-1}(l_{3}) & \cdots & P_{n-1}(l_{n})
\end{pmatrix},
\end{equation}
where $\{P_{j}(x)\}_{j=0}^\infty$ is the system of monic polynomials 
orthogonal  with respect to the weight $w(l)$.
 Then (\ref{dph7}) becomes 
\begin{equation} \label{app2}
\tau_n=\frac{2^{n^2}}{n!}\sum_{l_1,\ldots,l_n=1}^\infty
\left(\sum_{\pi\in S_{n}}(-1)^\pi\prod_{k=1}^{n}P_{\pi(k)-1}(l_{k})\right)^{2}
\prod_{k=1}^{n}w(l_{k}).
\end{equation}
Note that the orthogonality condition ensures that, after summing, only diagonal terms are non-zero, 
so we get 
\begin{equation} \label{app3}
\tau_{n}=\frac{2^{n^2}}{n!}\sum_{l_1,\ldots,l_n=1}^\infty\left(\sum_{\pi\in S_{n}}
\prod_{k=1}^{n}P_{\pi(k)-1}^{2}(l_{k})\right)
\prod_{k=1}^{n}w(l_{k})=2^{n^{2}}\prod_{k=0}^{n-1}h_{k}.
\end{equation}

\end{document}